\documentclass{PoS}
\usepackage{lineno}
\usepackage{epstopdf}
\title{Measurements of open-charm production in pp and p--Pb collisions with the ALICE detector at the LHC}

\ShortTitle{Measurements of open-charm production in pp and p--Pb collisions with the ALICE detector at the LHC}

\author{\speaker{Chitrasen Jena, for the ALICE Collaboration}%
         \thanks{Presently at School of Physical Sciences, National Institute of Science Education and Research, Jatni - 752050, India}\\
         Universit\`a  degli Studi di Padova and INFN Sezione di Padova\\
        E-mail: \email{jena@pd.infn.it}}

\abstract{ Hadrons containing heavy quarks, i.e. charm and beauty, are effective probes to investigate the properties of the hot, dense 
and strongly-interacting medium formed in high-energy nuclear collisions. The relatively large masses of heavy 
quarks ensure that they are predominantly produced in the early stages of the collision and probe the complete
space-time evolution of the expanding medium. The measurements of D-meson production in pp collisions provide 
an important test of pQCD calculations and serve as an essential baseline for the comprehensive studies in heavy-ion 
collisions. The study of D-meson production in p--Pb collisions is necessary to disentangle the cold nuclear matter effects 
from hot nuclear matter effects. The measurement of heavy-flavour production as a function of charged-particle multiplicity 
in pp and p--Pb collisions could provide insight into the role of multi-parton interactions at LHC energies. 

We present ALICE results on D-meson production in pp collisions at $\sqrt{s} =$ 7 TeV and p--Pb collisions at 
$\sqrt{s_{\rm {NN}}}=$ 5.02 TeV. The D-meson yields per event, measured in different multiplicity intervals and normalized 
to their multiplicity-integrated values, are presented for pp and p--Pb collisions. The $p_{\rm {T}}$-differential production 
cross section and nuclear modification factor of prompt D mesons are measured in p--Pb collisions. The nuclear modification 
factor, $R_\mathrm{pPb}$, is compatible with unity within uncertainties, indicating that cold nuclear matter effects are small for 
$p_{\mathrm{T}} \gtrsim 3$~GeV/$c$. The D-meson transverse momentum distributions in p--Pb collisions relative to pp collisions, measured 
in several multiplicity classes, are also discussed.}

\FullConference{ 7th International Conference on Physics and Astrophysics of Quark Gluon Plasma\\
                 1-5 February , 2015 \\
                 Kolkata, India}

\begin{document}

\section{Introduction}
The primary goal of the ALICE experiment at the LHC is to study the strongly-interacting matter created in 
high-energy heavy-ion collisions. This state of matter with deconfined quarks and gluons, called Quark-Gluon Plasma (QGP),
is predicted by Quantum Chromodynamics (QCD) to exist at high temperatures and/or high energy densities~\cite{qcd1}. Heavy quarks are a 
powerful probe for investigating the properties of the QGP, since they are predominantly produced in initial hard scattering 
processes and experience all the stages of medium evolution. The measurements of heavy-flavour production in pp collisions 
allow for precision tests of perturbative QCD (pQCD) calculations and provide an essential reference for understanding the 
results from heavy-ion collisions~\cite{alicepbpbraa2}. Several cold nuclear matter (CNM) effects, such as 
the modification of parton distribution functions and momentum broadening due to parton scattering in the nucleus, 
can affect heavy-quark production and cannot be accounted for based only on pp data~\cite{cnm1}. It is therefore 
necessary to study p--Pb collisions, where an extended hot and dense strongly-interacting medium is not expected to form, to quantify 
these CNM effects. In addition, heavy-flavour production as a function of charged-particle multiplicity in pp and p--Pb collisions is 
expected to be sensitive to the interplay between hard and soft QCD processes and could provide insight into the relevance of 
Multi-Parton Interactions (MPIs) on heavy-flavour production~\cite{PYTHIA8,percolation,EPOS3}. 

\section{D-meson reconstruction with ALICE}
The measurement of charm production in ALICE is performed by reconstructing 
$\mathrm{D}^0$, $\mathrm{D}^+$, $\mathrm{D}^{*+}$ and $\mathrm{D}_{\mathrm{s}}^+$ via their hadronic decays 
$\mathrm{D}^0\rightarrow{}\mathrm{K}^{-}\pi^+$, 
$\mathrm{D}^+\rightarrow{}\mathrm{K}^{-}\pi^+\pi^+$,
$\mathrm{D}^{*+}\rightarrow{}\mathrm{D}^0\pi^+$ and
$\mathrm{D}_{\mathrm{s}}^+\rightarrow\phi\pi^+\rightarrow\mathrm{K}^+\mathrm{K}^-\pi^+$ 
and their charge conjugates. The analysis strategy for the D mesons in the central rapidity region 
is based on the invariant mass analysis of fully reconstructed decay topologies originating from displaced vertices. 
The Inner Tracking System (ITS) detector provides high spatial resolution of the track impact parameter allowing the 
reconstruction of secondary vertices from heavy-flavour decays. 
The study of the decay topology allows for an efficient rejection of the combinatorial background from uncorrelated tracks. 
In order to further suppress the combinatorial background, the Time Projection Chamber (TPC) and the Time Of Flight (TOF) detectors are 
used to identify pions and kaons. Details of the analysis procedure in pp and p--Pb collisions are reported in
~\cite{ALICEpp7TeVDmesons} and~\cite{aliceRpPb}, respectively.

\section{D-meson production in pp collisions}
\begin{figure}[h!t]
\centering
\begin{minipage}{10.0pc}\vspace{-0.6cm}
\includegraphics[width=12.0pc]{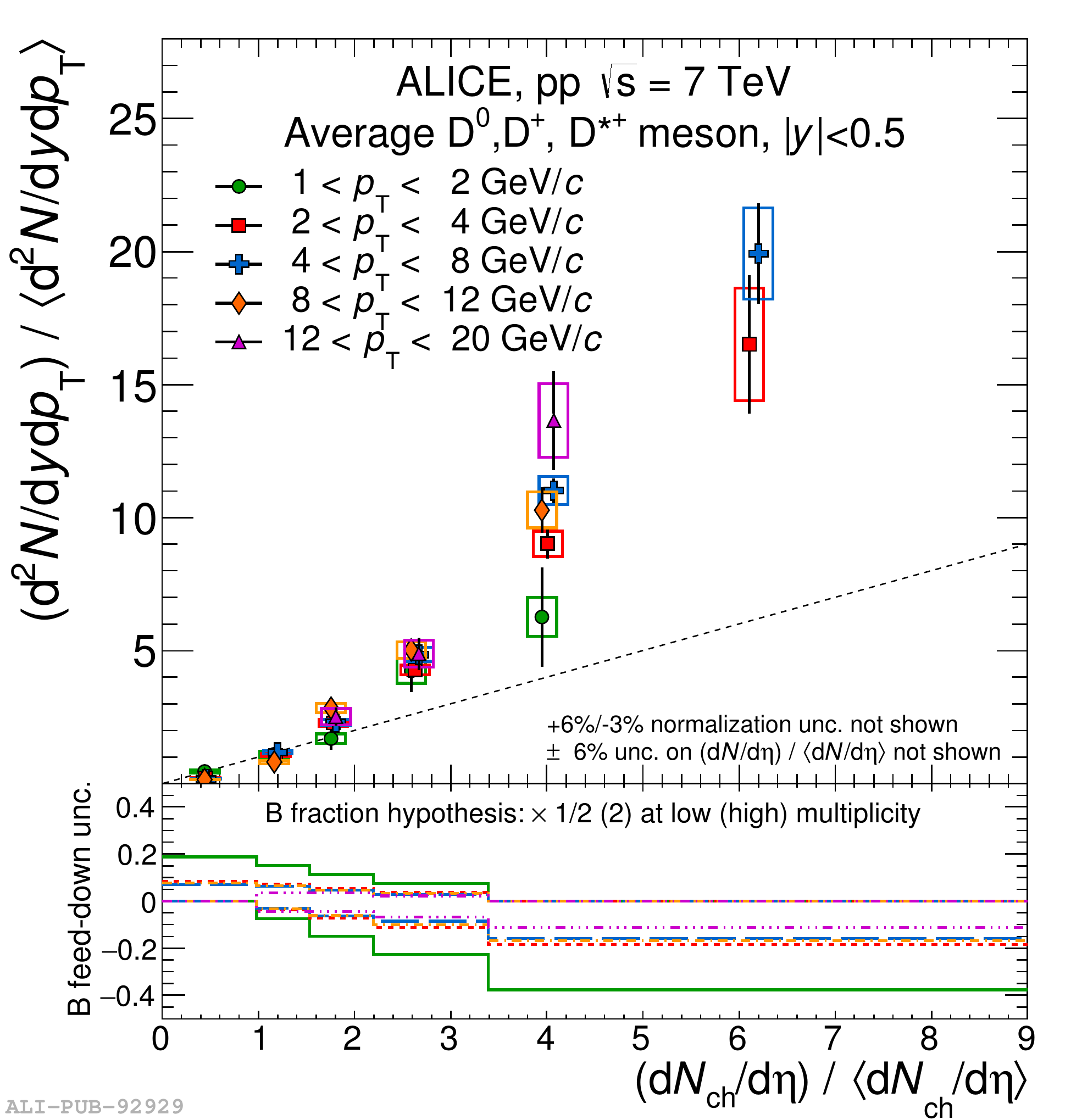}
\end{minipage}\hspace{2pc}%
\begin{minipage}{10.0pc}\vspace{-0.6cm}
\includegraphics[width=12.0pc]{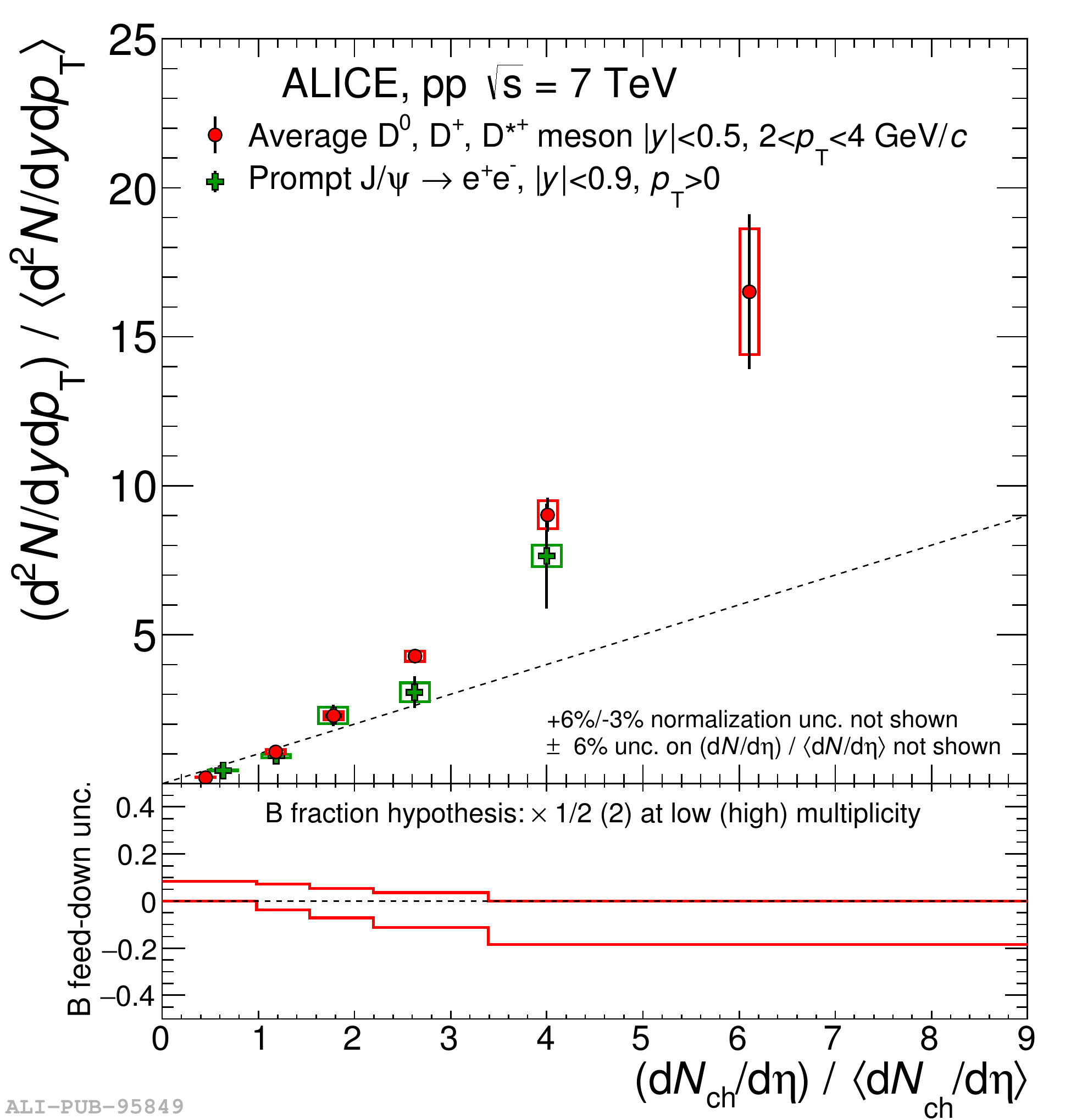}
\end{minipage}\hspace{2pc}%
\begin{minipage}{10.0pc}\vspace{-0.6cm}
\includegraphics[width=12.0pc]{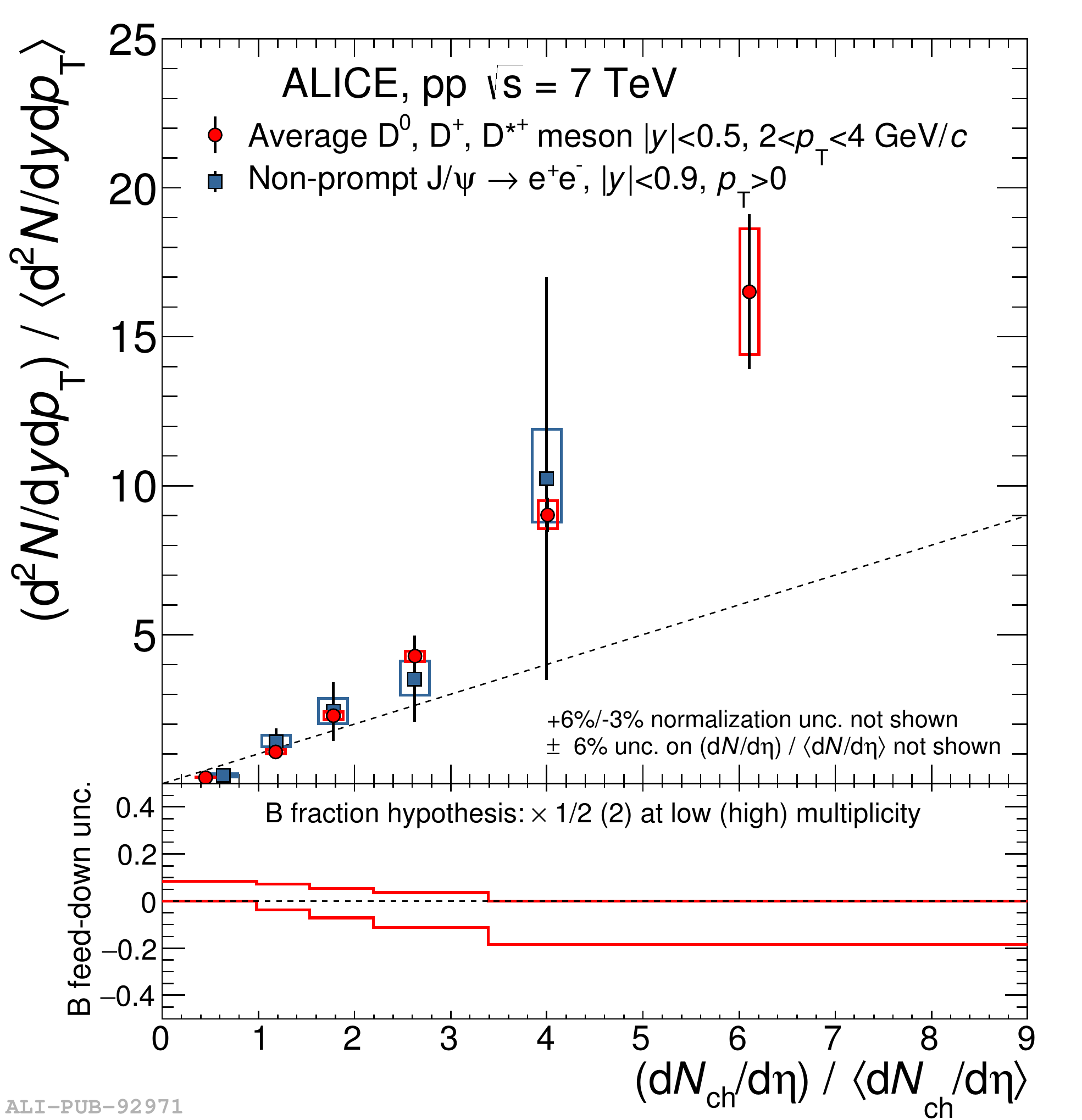}
\end{minipage} 
\caption{\label{fig:DmesonvsMult}\label{fig:DmesonvsMult} Average of D$^0$, D$^+$ and D$^{*+}$ relative yields as a function of the relative 
charged-particle multiplicity at central rapidity in different $p_\mathrm{T}$ intervals (left panel). Average prompt D meson relative yield as a function 
of the charged-particle multiplicity at central rapidity compared with the relative yields of prompt (middle panel) and non-prompt J/$\psi$ (right panel)~\cite{DmesonMultpp}.}
\end{figure}
The $p_\mathrm{T}$-differential production cross sections were measured for prompt D-mesons in pp collisions at 
$\sqrt{s}=7$ TeV~\cite{ALICEpp7TeVDmesons} and 2.76 TeV~\cite{ALICEpp276TeVDmesons}. The measured cross sections are 
well described by pQCD calculations~\cite{FONLL,GM-VFNS,kTfact}.

The study of D-meson production evaluated for various multiplicity and $p_\mathrm{T}$ intervals, is presented via the D-meson relative yields, i.e. 
the ratio of the yields in each multiplicity interval to the multiplicity-integrated (average) yield:
\begin{equation} \label{eq:relYield}
\frac{\left(\mathrm{d^{2}}N^{\mathrm{D}}/\mathrm{d}y\mathrm{d}p_{\mathrm{T}}\right)^{j}}{\langle \mathrm{d^{2}}N^{\mathrm{D}}/\mathrm{d}y\mathrm{d}p_{\mathrm{T}}\rangle}
= \left(\frac{1}{N^{j}_{\mathrm{events}}}\frac{N^{j}_{\mathrm{raw~D}}}{\epsilon^{j}_{\mathrm{prompt~D}}}\right)\Biggm/\left(\frac{1}{N_{\mathrm{MB~trigger}}/\epsilon_{\mathrm{MB~trigger}}}\frac{\langle N_{\mathrm{raw~D}} \rangle}{\langle \epsilon_{\mathrm{prompt~D}}\rangle}\right), 
\end{equation} 
where the index $j$ identifies the multiplicity interval, $N^{j}_{\mathrm{raw~D}}$ is the raw yield, $\epsilon^{j}_{\mathrm{prompt~D}}$ is the 
reconstruction and selection efficiency for prompt D mesons and $N^{j}_{\mathrm{events}}$ is the number of events analysed in each multiplicity 
interval. The number of events used for the normalisation of the multiplicity integrated yield are corrected for the fraction of inelastic collisions that are not
selected by the minimum-bias trigger, expressed as $N_{\mathrm{MB~trigger}}/\epsilon_{\mathrm{MB~trigger}}$. 
The relative yields of D$^0$, D$^+$ and D$^{*+}$ mesons are compatible in all $p_\mathrm{T}$ intervals within uncertainties~\cite{DmesonMultpp}. The left panel of
Fig.~\ref{fig:DmesonvsMult} shows the average of the D-meson relative yield as a function of the charged-particle multiplicity at central rapidity (${\mathrm{d}}N_{\mathrm{ch}}/\mathrm{d}\eta)$ for different $p_\mathrm{T}$ intervals. The relative D-meson yield increases with the charged-particle multiplicity by about a factor of 15 in the 
range between 0.5 and 6 times $\langle {\mathrm{d}}N_{\mathrm{ch}}/\mathrm{d}\eta\rangle$ and the enhancement is independent of $p_\mathrm{T}$ within the 
uncertainties. The relative yields of prompt D mesons compared with the relative yields of prompt and non-prompt J/$\psi$, where non-prompt J/$\psi$
are those coming from b-hadron decays, 
are shown in the middle and right panels of Fig.~\ref{fig:DmesonvsMult}. A similar increase of relative yields for charm (open and hidden) and beauty with 
charged-particle multiplicity suggests
that the increasing trend is most likely due to the c$\bar{\mathrm{c}}$ and b$\bar{\mathrm{b}}$ production processes, and is not significantly influenced by 
hadronisation. The yield enhancement as a function of the charged-particle multiplicity is qualitatively described by PYTHIA 8 calculations including 
MPI contributions~\cite{PYTHIA8}, a percolation model taking into account the influence of colour charge exchanges during the interaction~\cite{percolation} 
and the EPOS 3 event generator which provides a description of the initial conditions followed by a hydrodynamical evolution~\cite{EPOS3}. More precise 
measurements are needed in order to further constrain the models.

\section{D-meson production in p-Pb collisions}
\begin{figure}[h!t]
\centering
\begin{minipage}{13.2pc}\vspace{-0.6cm}
\includegraphics[width=13.2pc]{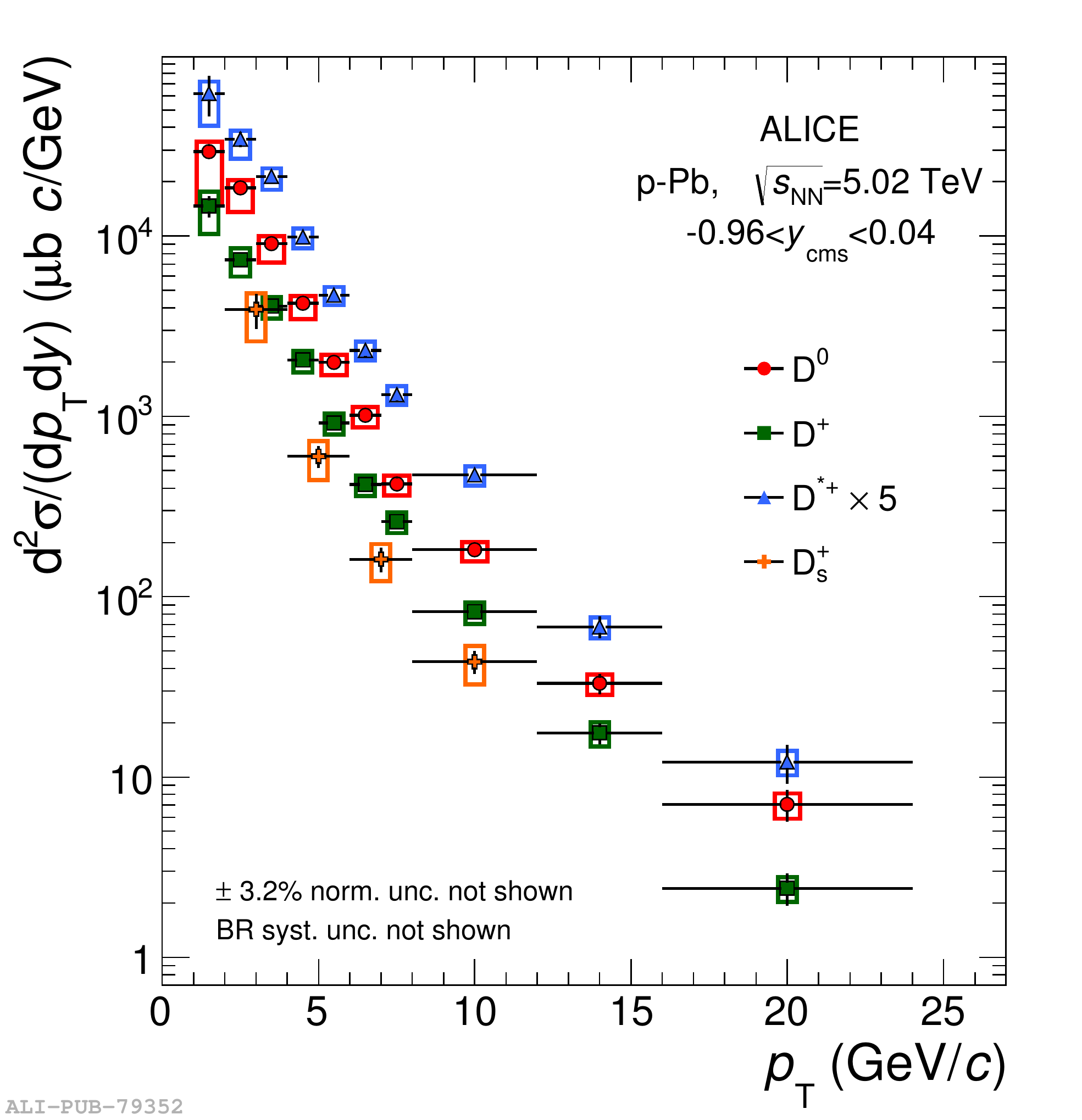}
\end{minipage}\hspace{2pc}%
\begin{minipage}{13.2pc}\vspace{-0.6cm}
\includegraphics[width=13.2pc]{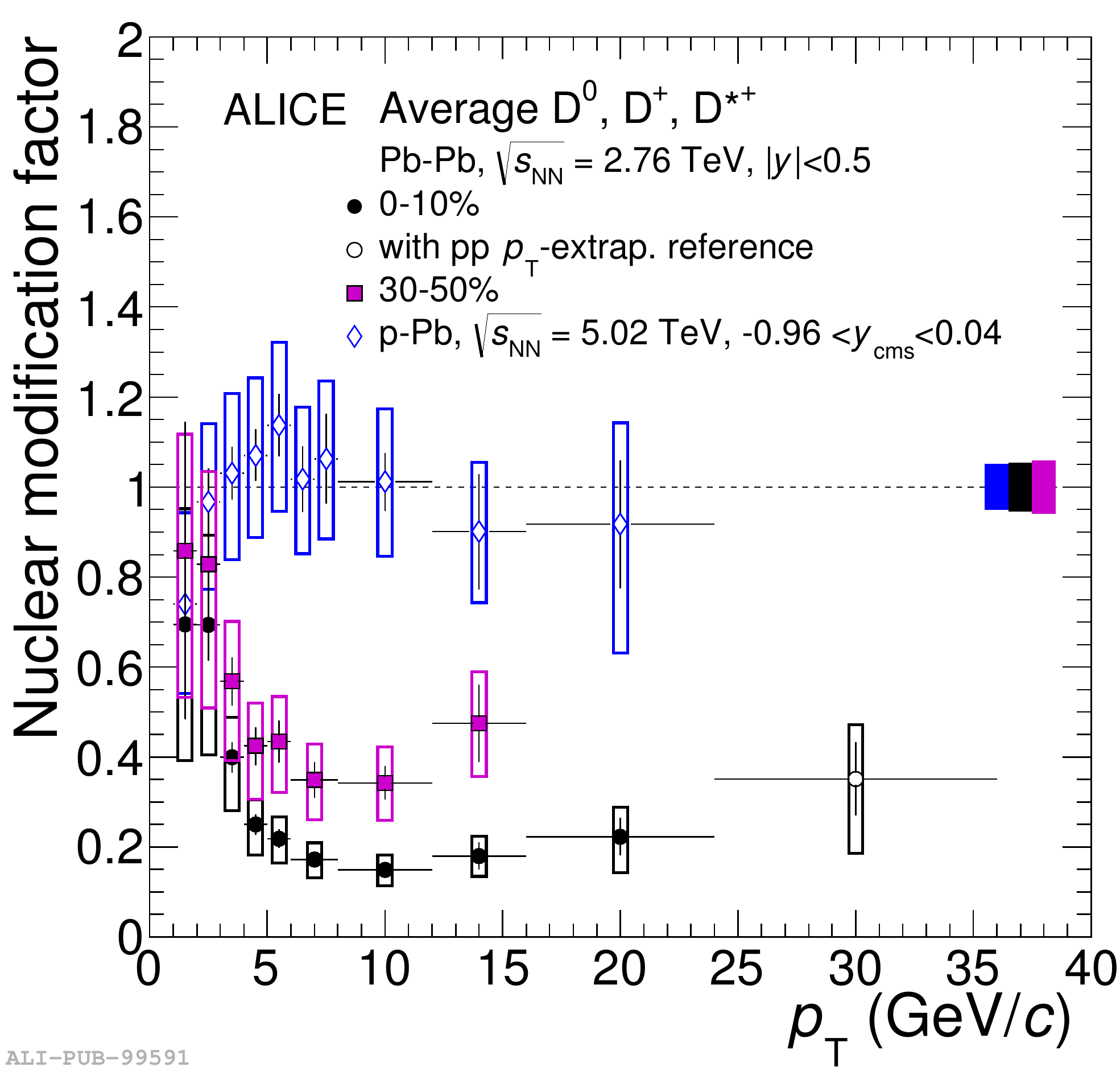}
\end{minipage} 
\caption{\label{fig:Dcrosssection}Left panel: $p_\mathrm{T}$-differential production cross section of prompt D$^0$, D$^+$, D$^{*+}$, and D$_\mathrm{s}^{+}$ 
mesons in p--Pb collisions at $\sqrt{s_\mathrm{NN}}=5.02$~TeV~\cite{aliceRpPb}. Right panel: Average $R_\mathrm{pPb}$ of D$^0$, D$^+$ and D$^{*+}$ mesons as 
a function of $p_\mathrm{T}$ in p--Pb collisions at $\sqrt{s_\mathrm{NN}}=5.02$~TeV compared to D-meson $R_\mathrm{AA}$ in central (0--10\%) and in 
semi-peripheral (30--50\%) Pb--Pb collisions at $\sqrt{s_\mathrm{NN}}=2.76$~TeV~\cite{alicepbpbraa2}.}
\vspace{-0.2cm}
\end{figure}
The left panel of Fig.~\ref{fig:Dcrosssection} shows the $p_\mathrm{T}$-differential production cross sections of prompt D$^0$, D$^+$, D$^{*+}$, and 
D$_\mathrm{s}^+$ mesons in p--Pb collisions at $\sqrt{s_\mathrm{NN}}=5.02$~TeV. The $p_\mathrm{T}$-differential cross sections were used to compute 
the nuclear modification factor, $R_\mathrm{pPb}$, defined as
$R_\mathrm{pPb}(p_\mathrm{T}) = \frac{\mathrm{d}\sigma_\mathrm{pPb}/\mathrm{d}p_\mathrm{T}}{A\times(\mathrm{d}\sigma_\mathrm{pp}/\mathrm{d}p_\mathrm{T})},$
where $\mathrm{d}\sigma_\mathrm{pPb}/\mathrm{d}p_\mathrm{T}$ is the $p_\mathrm{T}$-differential cross section in p--Pb collisions, 
$\mathrm{d}\sigma_\mathrm{pp}/\mathrm{d}p_\mathrm{T}$ is the $p_\mathrm{T}$-differential cross section in pp collisions at the same 
centre-of-mass energy, and $A$ is the mass number of the Pb nucleus. The reference pp cross sections at $\sqrt{s}=5.02$~TeV were 
obtained by a FONLL-based energy scaling of the $p_\mathrm{T}$-differential cross sections measured at $\sqrt{s}=7$~TeV~\cite{energyscaling}.
The right panel of Fig.~\ref{fig:Dcrosssection} shows the average $R_\mathrm{pPb}$ of prompt D mesons as a function of  $p_\mathrm{T}$ in p--Pb collisions 
at $\sqrt{s_\mathrm{NN}}=5.02$~TeV, compared to the average $R_\mathrm{AA}$ of prompt D meson in 
central (0--10\%) and semi-peripheral (30--50\%) Pb--Pb collisions at $\sqrt{s_\mathrm{NN}}=2.76$~TeV~\cite{alicepbpbraa2}. 
The $R_\mathrm{pPb}$ is close to unity within uncertainties, indicating that the suppression of D mesons for $p_{\mathrm{T}} \gtrsim 3$~GeV/$c$ 
observed in Pb-Pb collisions cannot be explained in terms of CNM effects but is due to strong final-state effects induced 
by hot and dense QCD matter.
\begin{figure}[h!t]
\centering
\begin{minipage}{13.2pc}\vspace{-0.2cm}
\includegraphics[width=13.2pc]{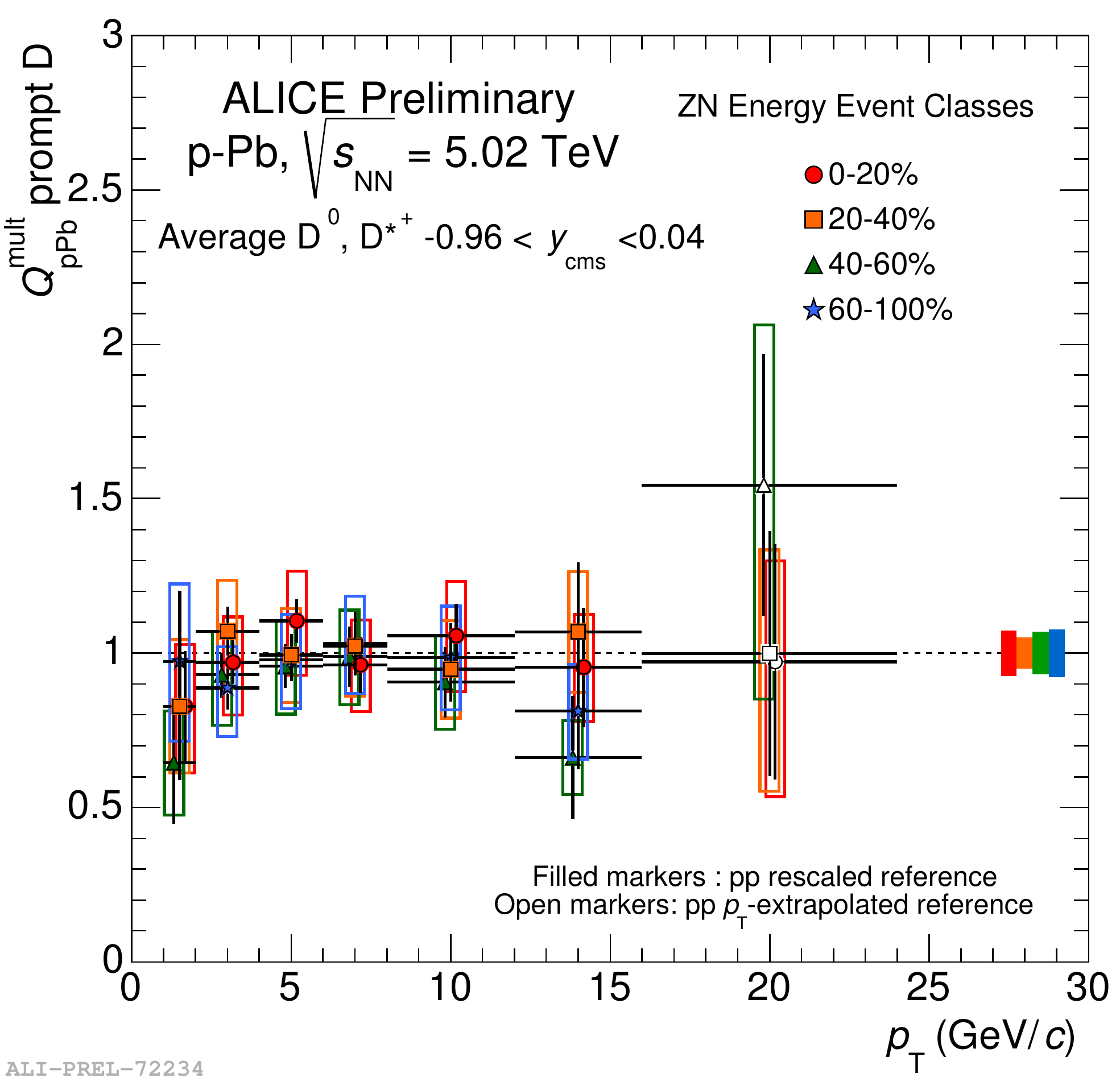}
\end{minipage}\hspace{2pc}%
\begin{minipage}{13.2pc}\vspace{-0.2cm}
\includegraphics[width=13.2pc]{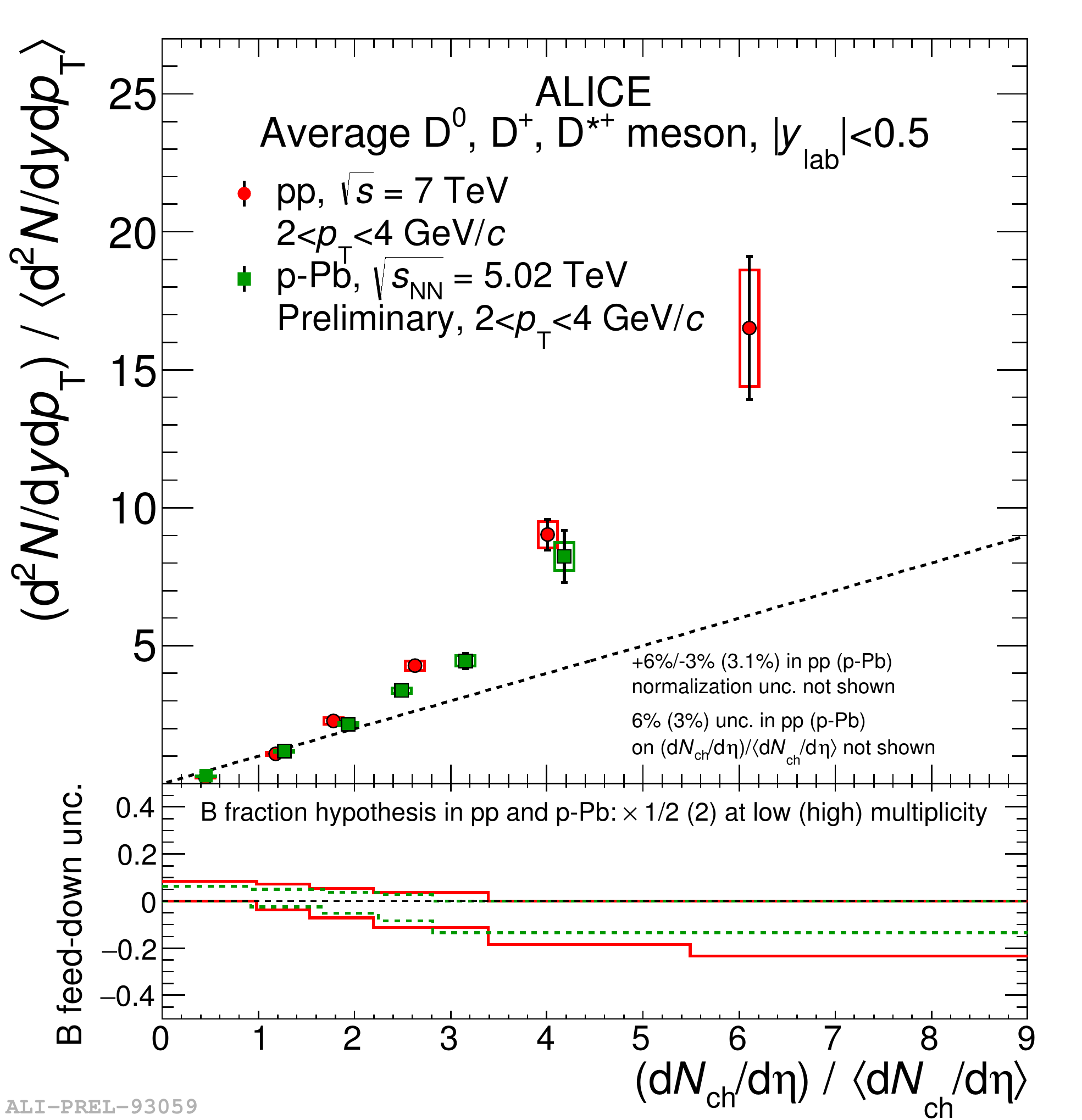}
\end{minipage} 
\caption{\label{fig:QpPbnMult}Left panel: Average D$^0$, D$^+$ and D$^{*+}$  nuclear modification factors as a function of $p_\mathrm{T}$  
in the 0--20\%, 20--40\%, 40--60\% and 60--100\% event classes selected with the ZNA estimator. Right panel: Relative D$^0$ 
yield as a function of multiplicity for $2 < p_\mathrm{T} < 12$~GeV/$c$ in p--Pb collisions, compared with the results from pp collisions.}
\end{figure}

The production of prompt D$^0$, D$^+$ and D$^{*+}$ mesons is also studied in four event classes using different centrality estimators. 
The nuclear modification factor is calculated as
$Q_\mathrm{pPb}(p_\mathrm{T}) = \frac{\mathrm{d}N^{\mathrm{cent}}_{\mathrm{pPb}}/\mathrm{d}p_\mathrm{T}}{\langle T^{\mathrm{cent}}_\mathrm{pPb}\rangle\times(\mathrm{d}\sigma_\mathrm{pp}/\mathrm{d}p_\mathrm{T})}$, 
where $\mathrm{d}N^{\mathrm{cent}}_\mathrm{pPb}/\mathrm{d}p_\mathrm{T}$ is the yield of prompt D mesons in p--Pb collisions in a given event class, 
$\mathrm{d}\sigma_\mathrm{pp}/\mathrm{d}p_\mathrm{T}$ is the $p_\mathrm{T}$-differential cross section of prompt D mesons in pp collisions, 
and $\langle T^{\mathrm{cent}}_\mathrm{pPb}\rangle$ is the average nuclear overlap function in a given centrality class. In contrast to the multiplicity-integrated 
$R_\mathrm{pPb}$,  $Q_\mathrm{pPb}$ is influenced by potential biases in the centrality estimation that are not related to nuclear effects~\cite{centpPb}. 
The least biased estimator is based on the energy deposited by nuclear fragments in the Zero Degree Neutron Calorimeter in the Pb-going direction (ZNA).
The D-meson $Q_\mathrm{pPb}$ estimated with the ZNA is consistent with unity within uncertainties and is independent of the event activity, 
as shown in the left panel of Fig.~\ref{fig:QpPbnMult}. The relative yields of D-mesons were also studied in p--Pb collisions as a function of 
charged-particle multiplicity and compared with the corresponding measurements in pp collisions, as shown in the right panel of Fig.~\ref{fig:QpPbnMult}. 
A similar relative increase of the D-meson yield with charged-particle multiplicity is observed in both pp and p--Pb collisions. It is worth noting that, in addition 
to the MPI, multiple binary nucleon-nucleon interactions occur in most p--Pb collision events and the collision initial conditions are also modified.

\section{Conclusions}
The measurements of D-meson production at central rapidity in pp and p--Pb collisions, performed with the ALICE detector, 
have been presented. The D-meson relative yields are found to increase with increasing charged-particle multiplicity in both pp and p--Pb collisions.
Charm (open and hidden) and beauty hadron relative yields exhibit a similar increase with increasing charged-particle multiplicity
at central rapidity in pp collisions at $\sqrt{s}=7$ TeV.  The $Q_{\mathrm{pPb}}$ of the D mesons measured in several event classes, 
defined with the least-biased estimator ZNA, are consistent with unity and also with the multiplicity-integrated $R_{\mathrm{pPb}}$. Therefore, from our data 
there is no evidence of a multiplicity dependence of D-meson production in p--Pb collisions with respect to that of pp collisions 
at the same centre-of-mass energy within the uncertainties.


\begin{thebibliography}{99}
\bibitem{qcd1} F. Karsch, Nucl. Phys. A  698 (2002) 199 and references therein.
\bibitem{alicepbpbraa2} J. Adam {\it et al.} (ALICE Collaboration), arXiv:1509.06888 [nucl-ex] and references therein.
\bibitem{cnm1} R. Vogt, arXiv:1508.01286 [hep--ph] and references therein.
\bibitem{PYTHIA8} T. Sjostrand, S. Mrenna, and P. Z. Skands, Comput. Phys. Commun. {\bf 178} (2008) 852.
\bibitem{percolation} E. G. Ferreiro and C. Pajares, Phys. Rev. C {\bf 86} (2012) 034903.
\bibitem{EPOS3} K. Werner, B. Guiot, I. Karpenko, and T. Pierog, Phys. Rev. C {\bf 89} (2014) 064903.
\bibitem{ALICEpp7TeVDmesons} B. Abelev {\it et al.} (ALICE Collaboration), JHEP {\bf 1201} (2012) 128.
\bibitem{aliceRpPb} B. Abelev {\it et al.} (ALICE Collaboration), Phys. Rev. Lett. {\bf 113} (2014) 232301.
\bibitem{ALICEpp276TeVDmesons} B. Abelev {\it et al.} (ALICE Collaboration), JHEP {\bf 1207} (2012) 191.
\bibitem{FONLL} M. Cacciari, M. Greco and P. Nason, JHEP {\bf 9805} (1998) 007.
\bibitem{GM-VFNS} B. A. Kniehl {\it et al.}, Eur. Phys. J. C {\bf 72} (2012) 2082.
\bibitem{kTfact} R. Maciula and A. Szczurek, Phys. Rev. D {\bf 87} (2013) 094022.
\bibitem{DmesonMultpp}  J. Adam {\it et al.} (ALICE Collaboration), JHEP {\bf 1509} (2015) 148. 
\bibitem{energyscaling} R. Averbeck {\it et al.}, arXiv:1107.3243 [hep--ph].
\bibitem{centpPb} J. Adam {\it et al.} (ALICE Collaboration), Phys. Rev. C {\bf 91} (2015) 064905.
\end{thebibliography}
\end{document}